\documentstyle[preprint,aps]{revtex}
\title{Aharonov-Bohm oscillations in a mesoscopic
ring with a quantum dot}
\author{A. Levy Yeyati\cite{aly} and M. B\"uttiker}
\address{D\'epartement de physique th\'eorique,
Universit\'e de Gen\`eve, \\ 4, Quai Ernest Ansermet,
CH-1211 Gen\`eve }
\begin{document}
\draft
\maketitle
\begin{abstract}
We present an analysis of the Aharonov-Bohm oscillations for
a mesoscopic ring with a quantum dot inserted in one of its arms.
It is shown that microreversibility demands that the phase of the
Aharonov-Bohm oscillations changes {\it abruptly} when
a resonant level crosses the Fermi energy. We use the Friedel sum
rule to discuss the conservation of the parity of the oscillations
at different conductance peaks. Our predictions are illustrated with
the help of a simple one channel model that permits the variation
of the potential landscape along the ring.
\end{abstract}

\vspace{1cm}

\narrowtext
A recent experiment by Yacoby et al. \cite{Yacoby}
investigated the Aharonov-Bohm (AB) oscillations in a ring with a
quantum dot (see Fig. 1). This experiment is of fundamental interest since
it depends not only on the total transmission through the quantum dot
but also on the phase accumulated by carriers traversing the dot.
The experiment thus gives a direct demonstration that coherent resonant
tunneling and sequential tunneling are not
equivalent\cite{MarkusI,Wingreen,vanHouten}.
Yacoby et al. emphasize two features of the Aharonov-Bohm oscillations:
First, it was found that the phase of the AB oscillations changes abruptly
whenever transmission through the quantum dot
reaches a peak. Second it was found that the AB oscillations at
consecutive conductance peaks are in phase. Here we discuss these
two observations, invoking only basic physical principles, and
illustrate them with a simple model calculation.

First, consider the phase jump of $\pi$ in the AB-oscillations, which is
observed each time a resonant condition is achieved.
In a two terminal conductance experiment the
measured conductance is necessarily an even function of the
AB flux through the ring\cite{Markus2,Markus3}, $G(\Phi) =  G(-\Phi)$.
In a Fourier
representation of the conductance

\begin{equation}
G(\Phi) = G_{0} + \Delta cos(2\pi \Phi/\Phi_{0} + \delta) + ...  ,
\end{equation}
this implies that the phase $\delta$ can only be either zero or $\pi$
but nothing in between. In the experiment the phase $\delta$ is a function of
gate voltage. If a phase change occurs as function of gate voltage it must,
therefore, be a sharp jump of zero width. We call the two possibilities
$\delta = 0$ and $\delta = \pi$ the parity of the AB-oscillations.
In contrast Yacoby et al. compare the sharp phase jump
with an analysis that leads to a broad
transition of the phase, and violates
the symmetry demanded by microreversibility.
This leads Yacoby et al. to argue
that a sharp phase jump is in contradiction with a non-interacting
electron-transport picture. Early work on the transmission
through one-channel loops does indeed show a symmetry breaking
term\cite{Gefen}.
Closer inspection of this result shows that the transmission probability is an
even function of flux \cite{Markus3,Markus2} in a two terminal geometry.
Below we show that the abrupt phase change is
a consequence of microreversibility only. It
is a phenomena that occurs independently of
whether interactions are significant or not.
Moreover the phase jump is abrupt even if there exists inelastic
scattering. We conclude that any deviations from a sharp jump must be
a consequence of fluctuations in the external control parameters.

We analyze the second feature, the conservation of parity
of the AB oscillations at consecutive peaks,
with the help of the Friedel sum rule, which remains valid in the
presence of electron-electron interactions \cite{Langer,Levy93}.
The Friedel sum rule relates the phase $\Delta \eta$ accumulated by a carrier
traversing a region $\Omega$ to the electronic charge in this volume.
The increment of phase and charge are related by
\begin{equation}
dQ = e d\eta /\pi .
\end{equation}
If the volume $\Omega$ is chosen to include only the quantum dot
then each addition of an electron to the dot requires an
increase of $\eta$ by  $\pi$ (see Fig. 2).
Associated with this phase jump there is a parity change of the
AB-oscillations at each conductance peak.
Consequently the
AB-oscillations at consecutive conductance peaks would not be in phase.
This is in contrast with the experimental observation of Ref. \cite{Yacoby}.
However, what counts is not the phase of the quantum dot alone.
The ring structure is connected to leads which are in turn connected
to reservoirs. As will be shown bellow, it is the phase accumulated in
the entire coherence volume
which counts. As a consequence, the relative parity on contiguous resonances
might change if the addition of an electronic charge
to the quantum dot is accompanied by the addition of a charge
$\alpha e$ to the leads of the ring. Over large distances,
the arms of the ring can be expected to remain in a charge neutral state.
The additional charge is
most likely accumulated at the barriers which separate the arms of the
ring from the quantum dot. The physical reason is that the gate used
to regulate the charge on the dot couples capacitively also to the gates
used to form the barriers between dot and ring. A strict conservation of
parity of the AB-oscillations occurs if the total charge $(1+\alpha)e$
added is zero or
an even multiple of 2e. Interestingly, because the phase observed
in the transmission coefficient can only be $0$ or $\pi$ a ``phase-locking"
occurs. Even if the additional charge $\alpha$ is not exactly an odd integer
the parity of the AB oscillations at a number of
consecutive conductance peaks will be the same.
We expect that the parity
of the AB-oscillations is conserved only over a limited number of peaks
and that this number depends on the geometry and
electrostatic properties of the sample.

In order to understand the behavior of the AB oscillations
in a device like that of Fig. 1a we start by
analyzing a single channel
noninteracting model. Our aim is to investigate both the influence of
inelastic scattering within the dot and of the effective potential landscape
along the ring.
We use a tight-binding representation
of the electron states (the corresponding lattice model is represented
in Fig. 1b) which allows for a qualitative description of any
potential profile. The effect of the magnetic flux $\Phi$ is taken into
account by a phase factor affecting the hopping matrix elements
$V_{i,j}$. We denote by $L$, $R$, $D$ and $F$ the left and
right leads, the arm with the dot and the free arm.
The effective electrostatic potential on the dot arm is parametrized by
the quantities $\epsilon_D$ (dot potential), $\epsilon_B$ (barriers
height) and $\epsilon_0$ (potential outside the dot) which are
schematically represented in Fig. 1b.
Inelastic scattering is simulated by a
third lead\cite{MarkusI,Datta,Pasta,Hershfield}(denoted by $I$)
coupled to the dot arm by a hopping element $V_I$.

The transmission properties of this model can be
easily obtained in terms of Green functions \cite{Pasta,Hershfield,us}.
In the absence of inelastic scattering ($V_I = 0$), the two terminal
conductance is proportional to the transmission coefficient $T_{LR}$,
which can be written in terms of the retarded Green functions as
\cite{Fisher-Lee}

\begin{equation}
T_{LR} = 4 V^2_L V^2_R |G_{0,N+1}(E_F)|^2 Im g_L(E_F) Im g_R(E_F) ,
\end{equation}

\noindent
where $g_{L,R}(E_F)$ denote the local Green functions on the
uncoupled leads at the Fermi energy and $V_{L,R}$ are the hopping
elements connecting the ring to the leads.
One can establish a correspondence between $2 V_L V_R \sqrt{Im g_L(E_F)
Im g_R(E_F)} G_{0,N+1}(E_F)$ and the elastic transmission amplitude
$t$ for this single channel case. The phase $\eta$ of $t$ is, therefore, equal
to that of $G_{0,N+1}(E_F)$.

Taking the case where $V_L = V_R = 0 $ as the
unperturbed case for which the isolated ring Green functions are denoted by
$g_{i,j}$, $G_{0,N+1}$ can be written as

\begin{equation}
G_{0,N+1} = \frac{g_{0,N+1}}{( 1 - g_{0,0} \Sigma_L) (1 - g_{N+1,N+1}
\Sigma_R) - g_{0,N+1} \Sigma_R g_{N+1,0} \Sigma_L} ,
\end{equation}

\noindent
where $\Sigma_{L,R} = V^2_{L,R} g_{L,R}$.
For a ring without inelastic scattering the functions $g_{i,j}$ behave
as $\exp{[i \phi (i-j)/(N+1)]} f_{i,j}(\phi)$ where $2 \phi = \pi \Phi/\Phi_0$
is the phase associated to the magnetic flux and $f_{i,j}$ is a real
even function of $\phi$.
The transmission coefficient, therefore, satisfies the symmetry relation
$T_{L,R}(\Phi) = T_{L,R}(-\Phi)$, which implies that
$\partial T_{LR}/ \partial \Phi \rfloor_{\Phi = 0} = 0$ in this limit.

In the presence of inelastic scattering the isolated ring Green
functions $g_{i,j}$ get an extra phase which depends on the distance
$|i-j|$ and $\partial T_{LR}/ \partial \Phi \rfloor_{\Phi = 0} = 0$ no
longer holds. Notice, however, that time reversal symmetry always
implies that $T_{LR} (\Phi) = T_{RL} (-\Phi)$ \cite{Markus2}.
We can analyze the flux dependence of the two terminal conductance
in this case by coupling the ring to the third lead.
The condition of no net current flow through this lead yields a
two terminal conductance proportional to the total transmission
probability, given by \cite{MarkusI}

\begin{equation}
T_{total} = T_{LR} + \frac{T_{LI} T_{IR}}{1 - R_{II}} ,
\end{equation}

\noindent
where $R_{II}$ is the reflection probability on the third lead. Taking
into account the property $\sum_{j} T_{ij} = 1 - R_{ii}$ one can easily
show that

\begin{equation}
T_{total} = 1 - R_{LL} + \frac{T_{LI} T_{IL}}{1 - R_{II}} ,
\end{equation}

\noindent
and therefore $T_{\alpha \beta}(\Phi) = T_{\beta \alpha}(-\Phi)$ implies that
$T_{total}$ is an even function of the magnetic flux. This simple
calculation shows that even in the presence of inelastic scattering the
only possible phases for the AB oscillations are $0$ and $\pi$ and thus
the transition from one to the other should {\it always} be abrupt.
The only effect of inelastic scattering is to reduce the amplitude of
the AB oscillations
by decreasing the direct elastic transmission $T_{LR}$.

Notice that this result is also true at finite temperatures: thermal
averaging can degrade the amplitude of the AB oscillations but cannot
introduce additional phases between $0$ and $\pi$.
The only possible sources of phase smearing in the experiments should be
traced to fluctuations in the gate voltages.

We can thus study the parity of the AB oscillations by computing
$\Delta_2 = \partial^2 T_{LR} / \partial \Phi^2 \rfloor_{\Phi = 0}$
which tells us whether $\delta = 0$ ($\Delta_2 <0$) or $\delta = \pi$
($\Delta_2 >0 $).
We now show how the parity change in the AB effect is related to the parity
effect of the isolated ring. It is well known \cite{parity} that, for
the case of spinless electrons, the
ring with an odd number of particles has a diamagnetic response whereas
for even number the response is paramagnetic. In a noninteracting model
the parity is determined mainly by the uppermost occupied state.
Near a resonant level, we can approximate $G_{0,N+1}$ as

\begin{equation}
G_{0,N+1} \sim \frac{\psi_{n_0} \psi^*_{n_{N+1}}}{(E_F - \epsilon_n -
\Delta_n) + i \Gamma_n} ,
\end{equation}

\noindent
where $\epsilon_n$ is the isolated ring eigenvalue closest to $E_F$,
$\psi_{n_j}$ denote the components of the corresponding wavefunction, and
$\Delta_n$ and $\Gamma_n$ are the real and imaginary parts of the
electron self-energy due to coupling with the leads ($\Delta_n + i
\Gamma_n = |\psi_{n_0}|^2 V^2_L g_L + |\psi_{n_{N+1}}|^2 V^2_R g_R$). The only
flux sensitive quantities in this expression are $\epsilon_n$ and
$\psi_{n_j}$. In particular, $\psi_{n_j}(\phi) = \exp{[i\phi j/(N+1)]}
\psi_{n_j}(0)$, and one has

\begin{eqnarray}
\frac{\partial G_{0,N+1}}{\partial \phi} \rfloor_{\phi = 0} & \sim & -i
G_{0,N+1} , \nonumber \\
\frac{\partial^2 G_{0,N+1}}{\partial \phi^2} \rfloor_{\phi = 0} & \sim
& G_{0,N+1} \left[ -1 + \frac{1}{(E_F - \epsilon_n -
\Delta_n) + i \Gamma_n} \left( \frac{\partial^2 \epsilon_n}{\partial \phi^2}
\rfloor_{\phi = 0} \right) \right] ,
\end{eqnarray}

\noindent
where we have used that $\epsilon_n(\phi) = \epsilon_n(-\phi)$.
The behavior of $\Delta_2$ near a resonance is thus given by:

\begin{equation}
\Delta_2 \sim T_{LR} \frac{E_F - \epsilon_n -\Delta_n}{(E_F - \epsilon_n
-\Delta_n)^2 + \Gamma_n^2} \left( \frac{\partial^2 \epsilon_n}{\partial
\phi^2}\rfloor_{\phi = 0} \right) .
\end{equation}

We see that when the resonance corresponds to a paramagnetic state of
the isolated ring (i.e. $\frac{\partial^2
\epsilon_n}{\partial\phi^2}\rfloor_{\phi = 0} < 0$) $\Delta_2$ changes
from positive to negative as the state crosses the Fermi energy, while
the opposite behavior is found when $\frac{\partial^2
\epsilon_n}{\partial\phi^2}\rfloor_{\phi = 0} >0 $.

Next, let us investigate
why the phase of the AB oscillations on
contiguous dot resonances appears to be the same.
Within the spinless electrons model and assuming that the effect of the
dot gate is to modify the value of $\epsilon_D$ alone, Eq. (9) predicts
that the AB oscillations on contiguous resonances should be {\it out}
of phase. This is illustrated in Fig. 2 where $T_{LR}(\Phi=0)$
is plotted as a function of $\epsilon_D$. The full and dotted lines
indicate the regions where $\Delta_2$ is positive or negative
respectively. We also show the
phase of the transmission amplitude which, as commented above, is
proportional to the electronic charge accumulated within the sample
as $\epsilon_0 - \epsilon_D$ increases. As can be observed, this rigid
model for the potential landscape variation leads to an increase in the
charge of one electron each time a resonance is crossed.

In a real situation one expects
the potential in the regions close to the
QD (not only within the dot) to vary as the gate voltage is modified.
This effect can be included in our model
by allowing $\epsilon_0$ to vary together with $\epsilon_D$.
Let us assume that this variation can be described by
$\delta \epsilon_0 = a \delta \epsilon_D/(\rho_0 \Delta E)$ where
$\rho_0$ is the mean density of states for the ring regions where
the potential equals $\epsilon_0$ and $\Delta E$ is the mean separation
between dot resonances.
The actual relationship between $\epsilon_0$ and $\epsilon_D$ should
depend on the mutual capacitances between the ring and the gate
electrodes.
The effect of this self-consistency condition is simply to add a fractional
charge
$\alpha e \sim ae$ to the ring between two resonances.
Note that $\alpha$ and $a$ are in general not equal since
the charge added depends on the actual density of states and not the average
density of states $\rho_0$.

Fig. 3 illustrates the effect of increasing the parameter $a$.
Notice that the calculated transmission exhibits now a varying
background in addition to the dot resonances, which reflects the level
structure of the ring \cite{comment}.
In case (a) the extra charge added to the system is $\alpha e \sim 0.30e$ per
cycle. It can be observed that an additional phase jump appears close to
the third resonance. Notice that the second and third resonances exhibit now
the same parity.
For increasing $a$ new phase jumps appear between resonances.
In this way, when $\alpha \sim 1$ (Fig. 3 b) several peaks with the
same parity may be found.

Since the phase $\delta$ of the AB oscillations can only be $0$ or $\pi$
it is not necessary to add exactly a multiple of $2e$ to find the same phase
at consecutive peaks.
Instead, the parity of the AB-oscillations at the $n$-th resonance
will be determined by the
integer multiple of charge $en_{eff}$ where $n_{eff}$ is the integer
that is closest to the charge $n(1+\alpha)$ added after $n$ cycles.
For $-1 \le \alpha \le -0.5,$ (if the ring
and dot remain approximately charge neutral) this will create a sequence
of effective charge states $en_{eff}$ with $n_{eff} =0$ for
a number of cycles $k$. The parity will change after the first $k$
cycles which add half
an electronic charge and cause the effective charge state to jump to
$en_{eff} = e.$ Hence for this case the number of parity conserving cycles
is $k (1+\alpha) = 1/2$ or $k = (1/2) (1+\alpha)^{-1}.$
For $0.5 \le \alpha \le 1$ (if we add nearly two electrons)
we will still obtain an effective charge sequence
$en_{eff}$ with $en_{eff}$
equal to an even multiple of $e$
but only for a finite sequence of cycles.
The parity will change after $k$ cycles for which
a deficit of half
an electronic charge occurs.
For this case the number of parity conserving cycles
is $k (|\alpha - 1|) = 1/2$ or $k = (1/2) (|\alpha - 1|)^{-1}.$
If $\alpha$ is in the interval $ - 0.5 < \alpha < 0.5,$ then
the parity will change at every peak except, occasionally, when
the effective charge state jumps by $2e$. For $\alpha$ in this
interval we can at most observe two consecutive peaks which are in phase.
Thus we find that it is possible to observe many consecutive conductance
peaks at which the parity of the AB-oscillations is conserved
if $\alpha \sim -1$ or if $\alpha \sim 1.$ Which of the
two cases, the approximate preservation of overall charge neutrality,
or the addition of nearly two electrons (or another even multiple)
per cycle is realized in the experiment cannot be answered without a detailed
determination of the relevant capacitance matrix for the structure.

We therefore conclude that
within the spinless electron model the conservation of
parity of the AB phase on contiguous resonances is indicating that
either zero or an {\it even}
number of electrons are added to the system per cycle.
We expect that the inclusion of spin degrees of freedom do not change
our conclusions: The charging energy of the dot will ensure that in
each cycle at most one electronic charge can be added to the dot.
The important conclusion of our analysis is that
the phase of AB oscillations
is not related to the dot charge {\it alone} but to the total charge of
the system. It is the charge of the ring and the dot that counts.

\acknowledgements

This work was supported by the
Swiss National Science Foundation.
One of us (A.L.Y.) also acknowledges support by the European
Community under contract No.CI1*CT93-0247.

\begin{figure}
\caption{(a) Schematic representation of a
mesoscopic ring threaded by a magnetic flux $\Phi$
with a quantum dot included in one of its arms.
(b) Lattice model for this system.}
\end{figure}

\begin{figure}
\caption{Transmission probability as a function of dot potential
$\epsilon_D$ for fixed potential on the rest of the ring (fixed
$\epsilon_0$). The full and dotted lines indicate the regions of
positive and negative parity respectively (see text). The dashed line
corresponds to the phase of the transmission amplitude.}
\end{figure}

\begin{figure}
\caption{Same as in Fig. 2 but allowing $\epsilon_0$ to vary together
with $\epsilon_D$. Case (a) corresponds to parameter $a \sim 0.3$
and case (b) to $a \sim 1$.}
\end{figure}


\begin{references}
\bibitem[*]{aly} Present address: Departamento de F\'\i sica de la
Materia Condensada C-XII, Facultad de Ciencias. Universidad Aut\'onoma
de Madrid, E-28049 Madrid, Spain.
\bibitem{Yacoby} A. Yacoby, M. Heiblum, D. Mahalu, and H. Shtrikman,
Phys. Rev. Lett. {\bf 74}, 4047 (1995).
\bibitem{MarkusI} M. B\"uttiker, IBM J. Res. Develop. {\bf 32}, 63 (1988).
\bibitem{Wingreen} N. Wingreen et al., Phys. Rev. B {\bf 40}, 11834 (1989).
\bibitem{vanHouten} H. van Houten, C.W.J. Beenakker and A.A.M. Staring,
in Single Charge Tunneling; Edited by H. Grabert and M.H. Devoret,
Plenum Press, New York (1991).
\bibitem{Markus2} M. B\"uttiker, IBM J. Res. Develop. {\bf 32}, 317 (1988).
\bibitem{Markus3} M. B\"uttiker, Y. Imry and M. Ya. Azbel, Phys. Rev.
{\bf A30}, 1982 (1984).
\bibitem{Gefen} Y. Gefen, Y. Imry and M. Ya. Azbel, Phys. Rev. Lett.
{\bf 52}, 139 (1983).
\bibitem{Langer} J.S. Langer  and V. Ambegaokar, Phys. Rev. {\bf 121},
1090 (1961).
\bibitem{Levy93} A. Levy Yeyati, A. Mart\'{\i}n-Rodero and F. Flores,
Phys. Rev. Lett. {\bf 71}, 2991 (1993).
\bibitem{Datta} S. Datta, Phys. Rev. B {\bf 40}, 8169 (1989).
\bibitem{Pasta} J.L. D'amato and H.M. Pastawski, Phys. Rev. B {\bf 41},
7411 (1990).
\bibitem{Hershfield} S. Hershfield, Phys. Rev. B {\bf 43}, 11586
(1991).
\bibitem{us} A. Levy Yeyati, Phys. Rev. B {\bf 45}, 14189 (1992).
\bibitem{Fisher-Lee} D.S. Fisher and P.A. Lee, Phys. Rev. B {\bf 23},
6851 (1981).
\bibitem{parity} D. Loss and P. Goldbart, Phys. Rev. {\bf B43}, 13762 (1991);
H.F. Cheung, Y. Gefen, E.K. Riedel and W.H. Shih, Phys.
Rev. B {\bf 37}, 6050 (1988).
\bibitem{comment} This feature, which is not observed experimentally,
disappears in the numerical calculations when the ring
level separation is much smaller than the self-energy introduced by the
leads.
\end{references}
\end{document}